\documentclass{ws-ijmpcs}
\usepackage{graphicx,epsfig,wrapfig}

\begin{document}
\markboth{S. Kumano}
{Spin Physics at J-PARC}
\catchline{}{}{}{}{}

\title{Spin Physics at J-PARC}

\author{S. Kumano}

\address{KEK Theory Center, Institute of Particle and Nuclear Studies, KEK \\
1-1, Ooho, Tsukuba, Ibaraki, 305-0801, Japan \\
J-PARC Branch, KEK Theory Center,
Institute of Particle and Nuclear Studies, KEK \\
and Theory Group, Particle and Nuclear Physics Division, 
J-PARC Center \\
203-1, Shirakata, Tokai, Ibaraki, 319-1106, Japan\\
shunzo.kumano@kek.jp}

\maketitle

\begin{history}
\received{Day Month Year}
\revised{Day Month Year}
\published{Day Month Year}
\end{history}

\begin{abstract}
Spin-physics projects at J-PARC are explained by including
future possibilities. J-PARC is the most-intense hadron-beam 
facility in the high-energy region above multi-GeV, 
and spin physics will be investigated by using secondary beams 
of kaons, pions, neutrinos, muons, and antiproton as well as 
the primary-beam proton.
In particle physics, spin topics are on 
muon $g-2$, muon and neutron electric dipole moments, and
time-reversal violation experiment in a kaon decay.
Here, we focus more on hadron-spin physics as for future projects.
For example, generalized parton distributions
(GPDs) could be investigated by using pion and proton beams,
whereas they are studied by the virtual Compton scattering
at lepton facilities.
The GPDs are key quantities for determining the three-dimensional
picture of hadrons and for finding the origin of the nucleon spin
including partonic orbital-angular-momentum contributions.
In addition, polarized parton distributions and various hadron
spin topics should be possible by using the high-momentum beamline.
The strangeness contribution to the nucleon spin
could be also investigated in principle with the neutrino beam 
with a near detector facility.
\end{abstract}

\ccode{PACS numbers:13.85.-t, 24.85.+p, 12.38.-t, 13.40.Em, 13.20.Eb}

\section{Introduction to J-PARC}	

Japan Proton Accelerator Research Complex (J-PARC)
is located at Tokai in Japan,\cite{j-parc} and
it is a joint facility between KEK
(High Energy Accelerator Research Organization)
and JAEA (Japan Atomic Energy Agency) for projects in wide fields
of science. KEK is in charge of the particle- and nuclear-physics
projects at the 50-GeV proton synchrotron.
J-PARC provides most intense proton beam 
in the energy region above multi-GeV. 
Nuclear and particle physics projects use secondary beams
such as kaons, pions, neutrinos, muons, and antiproton
as well as the primary 50-GeV proton beam. 
The J-PARC experiments have been started for neutrino
oscillations and strangeness hadron experiments.
In this report, we explain spin physics at J-PARC, mainly
on possibilities of hadron spin physics.

J-PARC could cover a wide range of spin projects
from fundamental particle physics to hadron spin physics. 
They include 
muon $g-2$, muon and neutron electric dipole moments, and
time-reversal violation experiment in a kaon decay
as particle-physics projects. In hadron physics,
there are possible projects for clarifying the origin of nucleon spin
and associated three-dimensional structure of the nucleon.

\vspace{0.25cm}
\noindent
{\bf J-PARC facility}

\begin{figure}[b]
\begin{minipage}{0.48\textwidth}
\includegraphics[width=1.00\textwidth]{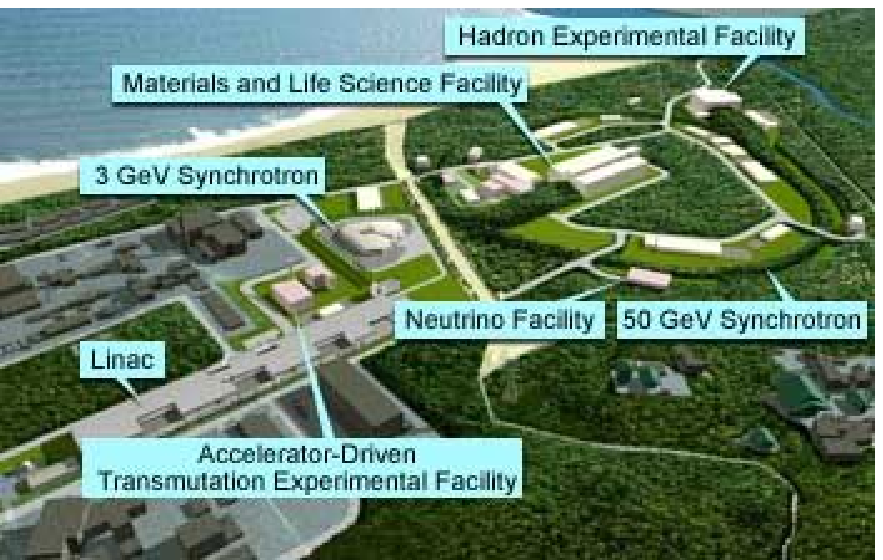}
\caption{J-PARC facility.$^1$}
\label{fig:j-parc}
\end{minipage}
\hspace{0.3cm}
\begin{minipage}{0.48\textwidth}
\includegraphics[width=1.00\textwidth]{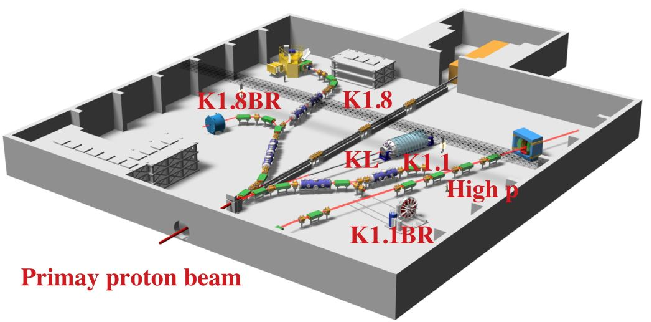}
\caption{Beamline layout of hadron hall.$^1$}
\label{fig:hadron-hall}
\end{minipage} 
\end{figure}

A bird's-eye view of the J-PARC facility is shown in Fig. \ref{fig:j-parc}.
The accelerator consists of a linac as an injector,
a 3-GeV rapid cycling synchrotron, and a 50-GeV synchrotron.
The energy of the 50-GeV synchrotron is 30 GeV at this stage.
J-PARC provides most intense proton beam in the high-energy region
($E>1$ GeV), and it is 1 MW in the 3-GeV synchrotron and 0.75 MW is expected
in the 50-GeV one. There are three major projects at J-PARC:
\begin{itemize}
\setlength{\leftskip}{-0.7cm}
\vspace{-0.25cm}
\item[$\bullet$]
material and life sciences as well as particle physics
with neutrons and muons produced by the 3-GeV proton beam,
\item[$\bullet$]
nuclear and particle physics with secondary beams (pions, kaons, 
neutrinos, muons, and antiprotons) by the 50-GeV proton beam and also with
protons of the 50-GeV primary beam,
\item[$\bullet$]
nuclear transmutation and neutron physics by the linac.
\end{itemize}
\vspace{-0.25cm}
Hadron-physics projects are investigated
at the hadron experimental facility in Fig.$\ $1.
Experiments on lepton-flavor violation and
time-reversal violation experiment in a kaon decay
will be done also in this hadron hall, in which
the beam-layout plan is shown in Fig. 2.
The K1.8 is intended to have kaons with momentum around
1.8 GeV/$c$ for the studies, for example,
on strangeness $-2$ hypernuclei with $\Xi^-$ by $(K^-,K^+)$ reactions.
The K1.1/0.8 beamline is designed for low-momentum stopped
kaon experiments such as the studies of kaonic nuclei.
The neutral kaon beamline (KL) is for studying CP violating
processes such as $K_L \rightarrow \pi^0 \nu\bar\nu$.
The ``High $p$" in Fig. 2 indicates
the high-momentum beamline for 50-GeV protons and unseparated
hadrons.
In the beginning stage of J-PARC, the proton-beam energy is
30 GeV instead of the original plan of 50 GeV. 
Measurements on muon $g-2$, muon and neutron electric dipole moment
will be done at the Materials and Life Science Experimental Facility (MLF)
and at the linac part for neutron physics.
Furthermore, hadron spin physics could be possible in principle
with the neutrino beam with a near detector, for example, by focusing on
strangeness spin in the axial form factor.

In future, hadron spin projects will become possible
at the hadron hall because the high-momentum beamline 
in Fig. 2 will be ready in the Japanese fiscal year of 2016.
The high-momentum beamline can transport the primary proton beam (30 GeV)
with the intensity of $10^{10}$-$10^{12}$/sec to the Hadron Hall, 
and it is a branch from the main proton beam of $10^{13}$ or 
$10^{14}$/sec.  In addition,
we can obtain unseparated secondary beams
such as pions etc. The beam intensity of these secondaries depends
on its species and momentum. A typical intensity for 10 GeV/c (15 GeV/c)
pions would be in the order of $10^7$/sec ($10^6$/sec). 
Because there is no approved proposal to study hadron spin physics
at J-PARC at this stage, we need to develop such possibilities.
In this article, we first introduce particle-spin projects,
and then we focus our discussions on high-energy hadron projects.

\vspace{-0.05cm}
\section{Spin in particle physics}	
\label{particle-spin}

Particle-spin projects are intended to probe physics beyond
the standard model. Here, we introduce three projects on
(1) muon $g-2$ and electric dipole moment,
(2) muon polarization in kaon decay, and
(3) electric dipole moment of the neutron.

\vspace{-0.15cm}
\subsection{Muon $g-2$ and electric dipole moment}

There is a project to measure the anomalous magnetic moment ($g-2$)
and electric dipole moment (EDM) of the muon
at the MLF of J-PARC.\cite{g-2-edm}
There is a long history of measurements on the muon $g-2$.
The muon magnetic moment is given by its spin and gyromagnetic ratio $g_\mu$
as $\vec \mu_\mu = g_\mu \frac{e}{2 m} \vec s$.
The anomalous magnetic moment $a_\mu$ is related to $g_\mu-2$ 
as $a_\mu = (g_\mu-2)/2$. We denote the muon EDM as $d_\mu$.
The electric dipole moment is given by the spin and $\eta_\mu$ as
$\vec d_\mu = \eta_\mu \frac{e\hbar}{2 m} \vec s$.
The most recent measurements were done by the BNL-E821 experiment,
and $a_\mu$ was measured down to 0.54 ppm 
and $d_\mu$ to 1.9$\times 10^{-19}$e$\cdot$cm.\cite{g-2-edm}
It is especially interesting that the current
value $a_\mu=0.00116592089(63)$
deviates from the theoretical value 
$0.00116591828(49)$ with 3.3$\sigma$ discrepancy.
This deviation exists even if theoretical uncertainties, 
mainly from hadronic corrections, are taken into account,
so that it could suggest new physics.
Therefore, it is important to improve the experimental measurement 
by an independent method.

\begin{wrapfigure}[8]{r}{0.40\textwidth}
   \vspace{-0.75cm}
\begin{center}
   \includegraphics[width=0.35\textwidth]{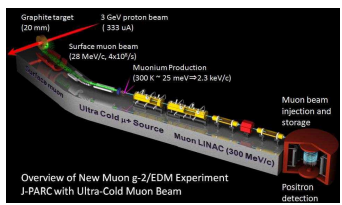}
\end{center}
    \vspace{-0.3cm}
\caption{$g-2$ and EDM experiment.$^2$}
\label{fig:g-2-edm}
\end{wrapfigure}

Under the static electric and magnetic fields, the muon spin 
precesses with the frequency 
\begin{align}
\! \! \!
\vec\omega 
=  - \frac{e}{m} 
    \left [ a_\mu \vec B - \left ( a_\mu - \frac{1}{\gamma^2-1} \right ) 
     \frac{\vec\beta \times \vec E}{c} \right ]
   - \frac{e}{m} 
    \left [ \frac{\eta}{2}
    \left ( \vec\beta \times \vec B + \frac{\vec E}{c} \right ) \right ]
\equiv \vec\omega_a + \vec\omega_\eta ,
\label{eqn:g-2-frequency-1}
\end{align}
where $\vec\omega_a$ is the precession vector due to
the anomalous magnetic moment, and $\vec\omega_\eta$
is the one due to the electric dipole moment.
In the CERN and BNL experiments, the muon energy was chosen 
to terminate the contribution of the $\vec\beta \times \vec E$ term,
and the EDM contribution $\vec\omega_\eta$ is neglected.
In the proposed J-PARC experiment in Fig.$\ $\ref{fig:g-2-edm},
the electric field $\vec E$ is terminated to give 
the precession frequency
\vspace{0.03cm}
\begin{align}
\vec\omega 
=  - \frac{e}{m} 
    \left [ a_\mu \vec B + \frac{\eta}{2} \, \vec\beta \times \vec B \right ] .
\label{eqn:g-2-frequency-2}
\vspace{0.03cm}
\end{align}
Here, the first $g-2$ term and the second EDM term are orthogonal
with each other, and they can be separated by appropriate
experimental design.
The purpose of this experiment is to improve the current limit
of $g-2$ to 0.1 ppm
by a factor of five
and the EDM to $10^{-21}$ e$\cdot$cm by a factor of two orders 
of magnitude than the current limit.
R\&D is in progress for significant technological developments 
to start this experiment within 2010's.\cite{g-2-edm}

\vspace{-0.20cm}
\subsection{Transverse muon polarization in kaon decay}

There is a project to investigate time-reversal violation
by the transverse muon polarization 
$P_T = \hat s_\mu \cdot (\hat p_{\mu^+} \times \hat p_{\pi^0} )$
in kaon decay $K^+ \to \pi^0 \mu^+ \nu$ at J-PARC.\cite{trek}
It is called TREK (Time-Reversal violation Experiment with Kaons)
experiment. This polarization could probe a signature beyond 
the standard model because the standard-model contributions 
from higher-order effects are considered to less than $10^{-6}$. 
The most recent measurement was done by the KEK-E246 experiment 
and it was
$P_T = - 0.0017 \pm 0.0023 \, (\text{stat}) \pm 0.0017 \, (\text{syst})$,
which is consistent with no T-violation.
New-physics models suggest that 
the polarization value should be as large as $10^{-3}$, 
which is just below the KEK-E246 limit.
Therefore, it is valuable to measure the polarization more accurately
to provide a clue for new physics.
In the proposed J-PARC experiment, background reduction will be
done by improved detector, especially the upgrade of polarimeter 
and magnet. The resulting error is expected to be $\delta P_T = 10^{-4}$.
In the beginning low-intensity period of J-PARC,
the collaboration proposed to start another experiment E36 on
the decay ratio 
$R_K = \Gamma (K^+ \to e^+ \nu) / \Gamma (K^+ \to \mu^+ \nu)$
to test lepton universality by using a sub-system of the TREK
experiment.\cite{trek}

\vspace{-0.20cm}
\subsection{Electric dipole moment of the neutron}

Ultracold neutrons (UCN) are used for measurements of fundamental
physics quantities, and new generation UCN sources are under
developments for experiments at facilities including 
J-PARC.\cite{ucn} In particular, the neutron electric dipole moment
(nEDM) should be able to probe new physics. 
For example, the supersymmetric model indicates that
the nEDM is of the order of $10^{-27}$ e$\cdot$cm, which is
just below the limit $3 \times 10^{-26}$ e$\cdot$cm
obtained by a Grenoble experiment. Since the limit
is constrained by the statistical error, efforts have been
made to increase the UCN density, as well as efforts
to reduce the systematic error.
Spallation UCN sources are
considered for the nEDM measurement at J-PARC.
Due to the increase of the proton-beam power to 20 kW,
the density is expected to become 
$10^3$-$10^4$ UCN/cm$^3$, 
whereas it was 0.7 UCN/cm$^3$ in the Grenoble experiment.
With the new UCN sources together with the intense proton beam,
much improvement will be made for the upper bound of the nEDM
into the region of $10^{-28}$ e$\cdot$cm. 
At J-PARC, the nEDM experiment is considered 
at the linac section.\cite{ucn}

\vspace{-0.27cm}
\section{Spin in hadron physics}	
\label{hadron-spin}

\vspace{-0.05cm}
\subsection{Flavor dependence of antiquark distributions}
\label{antiquark}
\vspace{-0.03cm}

\begin{wrapfigure}[10]{r}{0.41\textwidth}
   \vspace{-0.8cm}
\begin{center}
   \includegraphics[width=0.30\textwidth]{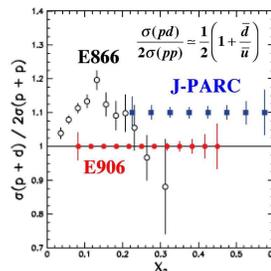}
\end{center}
    \vspace{-0.5cm}
\caption{Drell-Yan cross-section ratios.$^5$}
\label{fig:ub-db}
\end{wrapfigure}

There are proposals on dimuon experiments by using primary proton
beam for measuring parton distribution functions
(PDFs) in the medium Bjorken-$x$ region.\cite{p04-24} 
A typical example is shown in Fig. \ref{fig:ub-db} for 
measuring the light-antiquark distribution ratio
$\bar d (x)/\bar u(x)$ by the Drell-Yan processes of 
$pp$ and $pd$, where $p$ ($d$) is proton (deuteron).
According to perturbative QCD, the $\bar u/\bar d$ asymmetry
should be very small. Experimentally, the difference was suggested
by the violation of the Gottfried sum rule,\cite{flavor}
and it was confirmed explicitly by the Fermilab-E866 measurements
in Fig. \ref{fig:ub-db}. However, there is a peculiar tendency
of $\bar d/\bar u <1$ as $x$ becomes larger, which is difficult
to be understood theoretically.
The E906/SeaQuest experiment at Fermilab is in progress for dimuon
experiments to test this tendency, and it could be continued
at J-PARC.

Measurements of antiquark distributions in nuclei are also
interesting. The Fermilab Drell-Yan measurements showed
that nuclear modifications of the antiquark distributions are
very small at $x \sim 0.1$, which is in contradiction to 
the conventional pion-excess contribution.
It is important to confirm this result by an independent
experiment and to extend the measured $x$ region
for finding a physics mechanism of nuclear medium effects
on the antiquark distributions. In addition,
determination of parton distribution functions (PDFs)
at large $x$ is valuable for precisely calculating
other high-energy reactions, for example, high-$p_T$ jet and 
hadron production cross sections at LHC.

\vspace{-0.30cm}
\subsection{Generalized parton distributions}	
\vspace{-0.03cm}

\begin{wrapfigure}[7]{r}{0.36\textwidth}
\vspace{-0.80cm}
\begin{center}
\includegraphics[width=0.36\textwidth]{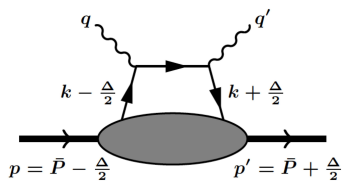}
\end{center}
\vspace{-0.5cm}
\caption{Kinematics for GPDs.}
\label{fig:gpd-1}
\end{wrapfigure}

Generalized parton distributions (GPDs) are key quantities for
studying three-dimensional structure of hadrons, and hence
to clarify the origin of the nucleon spin including partonic
orbital-angular-momentum contributions.\cite{gpd-summary}
They are measured typically in deeply virtual Compton scattering (DVCS)
as shown in Fig. \ref{fig:gpd-1}.
However, there are possibilities to investigate them at hadron
facilities like J-PARC.

First, we define kinematical variables for the process
$\gamma^* + p \to \gamma + p$ by the momenta
given in Fig. \ref{fig:gpd-1}. 
The average nucleon and photon momenta ($\bar P$ and $\bar q$)
and the momentum transfer $\Delta$ are defined by
$\bar P = (p+p')/2$, 
$\bar q = (q+q')/2$,
$\Delta = p'-p = q-q'$. 
Then, $Q^2$ and $t$ are given by
$Q^2 = -q^2$, $\bar Q^2 = - \bar q^2$, $t = \Delta^2$,
and the generalized scaling variable $x$ and a skewdness 
parameter $\xi$ are defined by
$x = Q^2 /(2p \cdot q)$, 
$\xi = \bar Q^2 /(2 \bar P \cdot \bar q)$.
The variable $x$ indicates the lightcone momentum fraction 
carried by a quark in the nucleon.
The skewdness parameter $\xi$ or the momentum $\Delta$ indicates
the momentum transfer from the initial nucleon to the final one
or the one between the quarks. 
The GPDs for the nucleon are given by off-forward matrix elements
of quark and gluon operators with a lightcone separation 
between nucleonic states.
The quark GPDs are defined by 
\begin{align}
 & \! \! \! \! \! \! \! \! 
 \int\frac{d y^-}{4\pi}e^{i x \bar P^+ y^-}
 \left< p' \left| 
 \overline{\psi}(-y/2) \gamma^+ \psi(y/2) 
 \right| p \right> \Big |_{y^+ = \vec y_\perp =0}
\nonumber \\
 & \! \! \! \! \! \! \! \! \! \! \! \!
 = \frac{1}{2  \bar P^+} \, \overline{u} (p') 
 \left [ H_q (x,\xi,t) \gamma^+
     + E_q (x,\xi,t)  \frac{i \sigma^{+ \alpha} \Delta_\alpha}{2 \, M}
 \right ] u (p) ,
\label{eqn:gpd-n}
\end{align}
where $H_q (x,\xi,t)$ and $E_q (x,\xi,t)$ are the unpolarized 
GPDs for the nucleon.

The major properties of the GPDs are the following.
In the forward limit ($\Delta,\, \xi,\, t \rightarrow 0$),
the nucleonic GPDs $H_q (x,\xi,t)$ become usual PDFs:
$H_q (x, 0, 0) = q(x)$.
Next, their first moments are the Dirac and Pauli form factors 
of the nucleon:
$\int_{-1}^{1} dx H_q(x,\xi,t)  = F_1 (t)$, 
$\int_{-1}^{1} dx E_q(x,\xi,t)  = F_2 (t)$.
Furthermore, the second moment is related to 
the quark orbital-angular-momentum contribution ($L_q$) to the nucleon spin:
$ J_q = \frac{1}{2} \int dx \, x \, [ H_q (x,\xi,t=0) +E_q (x,\xi,t=0) ]
      = \frac{1}{2} \Delta q + L_q$.
From this relation, we expect that the origin
of nucleon spin will be clarified by including
the orbital-angular-momentum contributions.
The GPDs contain information on the longitudinal 
momentum distributions and transverse structure 
as the form factors, so that they are appropriate quantities
for understanding the three-dimensional structure of hadrons.
The GPDs can be measured by the virtual Compton scattering
in Fig. \ref{fig:gpd-1} at lepton facilities. However, 
they can be measured also 
at hadron facilities such as J-PARC and GSI-FAIR, 
and examples are explained in the following subsections.

\vspace{-0.20cm}
\subsubsection{Exclusive Drell-Yan for studying GPDs}	

\begin{wrapfigure}[8]{r}{0.38\textwidth}
\vspace{-1.15cm}
\begin{center}
\includegraphics[width=0.36\textwidth]{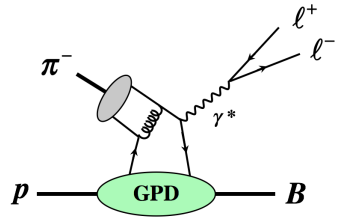}
\end{center}
\vspace{-0.4cm}
\caption{Exclusive Drell-Yan.}
\label{fig:ex-dy}
\end{wrapfigure}

The unseparated hadron beam, which is essentially the pion beam,
will become available at the high-momentum beamline of J-PARC,
so the exclusive Drell-Yan process is a possible future 
project at J-PARC. In Fig. \ref{fig:ex-dy}, the exclusive dimuon
process $\pi^- + p \to \mu^+ \mu^- + B$ is shown.\cite{ex-dy}
The process probes the nucleonic GPDs if $B$ is the nucleon, 
whereas it is related to the transition GPDs for $B \ne N$.
The cross section contains not only the GPDs but also the pion
distribution amplitude. Although there are models for this quantity
such as the asymptotic form or the Chernyak-Zhitnitsky type,
the pion part is rather well studied and tested experimentally
by the $\gamma \pi$ transition of Belle and BaBar. 
Using such studies for constraining the pion distribution amplitude, 
we should be able to obtain information on the GPDs by the exclusive
Drell-Yan process.

\vspace{-0.25cm}
\subsubsection{GPDs in the ERBL region}	

\begin{figure}[h!]
\begin{minipage}{0.405\textwidth}
   \vspace{-0.1cm}
\begin{center}
   \includegraphics[width=0.85\textwidth]{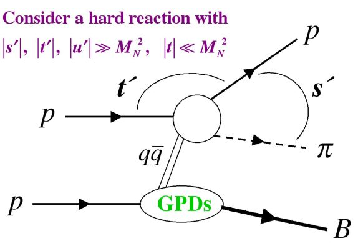}
\end{center}
   \vspace{-0.6cm}
\caption{GPD studies in $2 \to 3$ process.$^9$}
\label{fig:gpd-hadron}
\end{minipage}
\hspace{-0.10cm}
\begin{minipage}{0.59\textwidth}
   \vspace{0.15cm}
\begin{center}
   \includegraphics[width=1.00\textwidth]{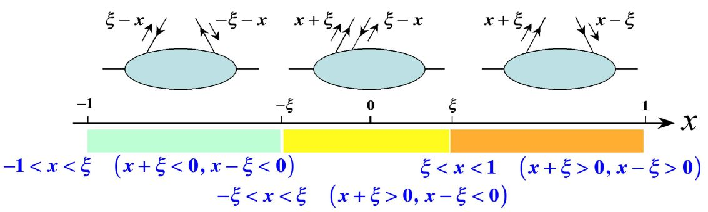}
\end{center}
\vspace{-0.3cm}
\caption{Three kinematical regions of GPDs.}
\label{fig:gpd-kinematics}
\end{minipage} 
\end{figure}

Using the high-energy proton beam, we could investigate the GPDs
at hadron facilities by using exclusive hadron-production
reactions $a+b \rightarrow c+d+e$ such as 
$N+N \rightarrow N + \pi + B$, as shown 
in Fig. \ref{fig:gpd-hadron}.\cite{gpd-hadron}
The GPD has three kinematical regions,
(1) $-1 < x < -\xi$, (2) $-\xi < x < \xi$, (3) $\xi < x < 1$,
as shown in Fig. \ref{fig:gpd-kinematics}.
The intermediate region (2) indicates
an emission of quark with momentum fraction $x+\xi$ with
an emission of antiquark with momentum fraction $\xi -x$.
The regions (1) and (3) are called as
DGLAP (Dokshitzer-Gribov-Lipatov-Altarelli-Parisi) regions, and
(2) is called the ERBL (Efremov-Radyushkin-Brodsky-Lepage) region. 
If the hadrons $c$ and $d$ have large and nearly opposite
transverse momenta and a large invariant energy, so that
an intermediate exchange could be considered as a $q\bar q$ state. 
Then, the $q\bar q$ attached to the nucleon 
is expressed by the GPDs in a special kinematical region of the ERBL
which is in the middle of three regions of Fig. \ref{fig:gpd-kinematics}.
This method of measuring the GPDs is complementary to 
the virtual Compton scattering in the sense that the specific
ERBL region is investigated in the medium $x$ region
due to the J-PARC energy of 30 GeV.

\vspace{-0.5cm}

\vspace{-0.00cm}
\subsection{Transverse-momentum-dependent distributions}

In order to understand the origin of nucleon spin including
orbital angular momenta, it becomes necessary to understand
the three-dimensional structure of the nucleon.
The momentum distribution along the motion of the nucleon is
the longitudinal parton distribution function, and the transverse
distribution plays a role of form factor at given $x$.
The inclusive Drell-Yan ($p+p\to \mu^+ \mu^- +X$) measurements
can probe transverse-momentum-dependent (TMD) polarized 
parton distributions, so called Boer-Mulders (BM) functions, 
by observing violation of the Lam-Tung relation in the cross 
sections.\cite{p04-24}
The BM functions indicate transversely-polarized quark distributions
in the unpolarized nucleon. 

\begin{figure}[h!]
  \vspace{-0.3cm}
\begin{center}
  \includegraphics[width=0.65\textwidth]{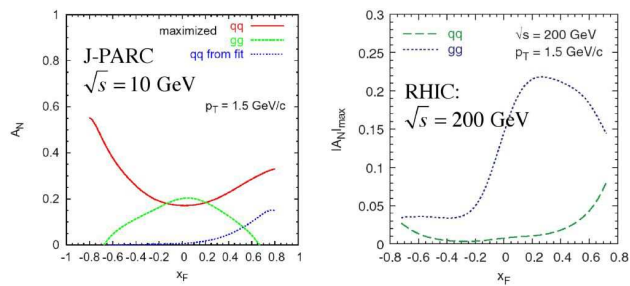}
\end{center}
     \vspace{-0.55cm}
\caption{Single spin asymmetry in $D$-meson production
 at J-PARC and RHIC.$^{5,11}$}
\label{fig:ssa-d}
\end{figure}

Other interesting TMD distributions are the Sivers functions,
which indicate unpolarized quark distributions in 
the transversely-polarized nucleon. For example, they appear
in single spin asymmetries of hadron-production processes
$p+ \vec p \rightarrow h +X$.
Their measurements are valuable not only for understanding transverse
structure but also for finding a relation to semi-inclusive
lepton measurement because the TMD distributions change sign
due to a difference of the gauge link.
It is analogous to the Aharonov-Bohm effect.\cite{ab-effect}
In order to show the advantage of J-PARC measurements,
we show single spin asymmetries
of $D$-meson production at J-PARC and RHIC
by considering the Sivers' mechanism in 
Fig. \ref{fig:ssa-d}.\cite{p04-24,j-parc-d}
The figures indicate the advantage that the quark (gluon) Sivers
functions are determined well at J-PARC (RHIC).

\vspace{-0.6cm}
\subsubsection{Tensor-polarized parton distribution functions}

\begin{wrapfigure}[10]{r}{0.42\textwidth}
   \vspace{-0.6cm}
\begin{center}
   \includegraphics[width=0.40\textwidth]{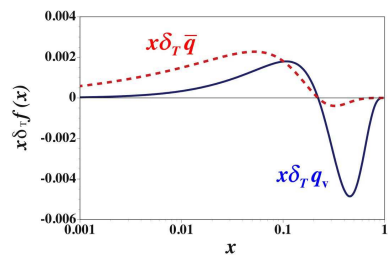}
\end{center}
    \vspace{-0.45cm}
\caption{Tensor-polarized PDFs.$^{12}$}
\label{fig:tensor-b1}
\end{wrapfigure}

In spin-one hadrons and nuclei such as the deuteron, there
exist new structure functions in charged-lepton deep inelastic
scattering, and they are called $b_1$, $b_2$, $b_3$, and $b_4$.
The twist-2 structure functions are $b_1$ and $b_2$, and they
are expressed in terms of the tensor-polarized quark 
distributions $\delta_T q (x)$.
The $b_1$ was first measured by the HERMES collaboration; however,
the detail of $x$ dependence is not clear and there is no accurate
data at medium and large $x$. The JLab $b_1$ proposal was approved
and its measurement will start in a few years.
The HERMES data indicated the violation of the sum rule 
$\int dx b_1(x)=0$,\cite{sk-b1} which suggests the existence
of finite tensor polarization in antiquark distributions.
This finite $\delta_T \bar q(x)$ can be directly
measured at hadron facilities by Drell-Yan processes 
with tensor-polarized deuteron 
($p + \vec d \rightarrow \mu^+ \mu^- +X$).\cite{sk-b1}
We note that the polarized-proton beam is not
needed for this experiment. This experiment is complementary
to the HERMES and JLab measurements of $b_1$
in the sense that the antiquark tensor polarization can be
measured specifically. The experiment will clarify
an exotic dynamical aspect of the deuteron in terms of
quark and gluon degrees of freedom, which cannot be done
in low-energy measurements.

\vfill\eject

\subsubsection{Elastic single-spin asymmetry}

\begin{wrapfigure}[12]{r}{0.30\textwidth}
   \vspace{-0.85cm}
\begin{center}
   \includegraphics[width=0.30\textwidth]{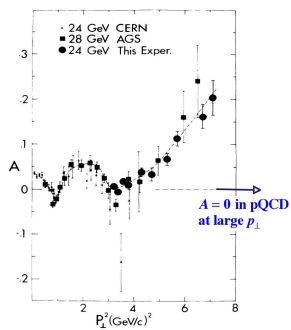}
\end{center}
    \vspace{-0.4cm}
\caption{Elastic single spin asymmetry in $p \vec p$.$^{13}$}
\label{fig:elasic-spin}
\end{wrapfigure}

The origin of the nucleon spin is one of unsolved issues in hadron physics;
however, there is another mysterious experimental result in elastic spin 
symmetries. For example, measurements indicated that 
the single-spin asymmetry in $p \vec p$ increases
as $p_\perp$ becomes larger at AGS in 
Fig. \ref{fig:elasic-spin}.\cite{elastic-spin}
According to perturbative QCD, the asymmetry has to vanish
at $p_\perp \to \infty$. Because the data are taken up to 
$p_\perp^2 =8$ GeV$^2$, nonperturbative physics might have contributed
to the finite asymmetry. Considering the peculiar feature of the AGS data,
we need to confirm the experimental measurements by an independent facility 
such as J-PARC before discussing possible physics mechanism.
However, rather than a mere confirmation, innovative methods 
should be developed, such as by changing targets
and by considering angular distributions in order to provide
clear evidences for theorists to understand the mechanism.

\vspace{-0.2cm}
\subsection{Spin physics with proton-beam polarization}\label{spin-pol}

\begin{wrapfigure}[10]{r}{0.45\textwidth}
   \vspace{-1.15cm}
\begin{center}
   \includegraphics[width=0.47\textwidth]{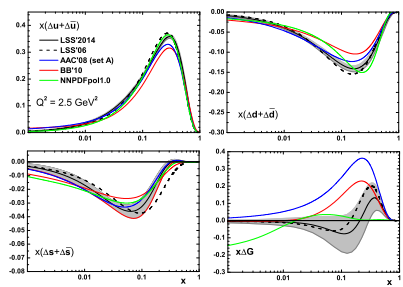}
\end{center}
    \vspace{-0.85cm}
\caption{Situation of polarized PDFs.$^{14}$}
\label{fig:polpdfs}
\end{wrapfigure}

It is technically possible to polarize the primary proton beam
at J-PARC.\cite{p04-24} However, it requires a major update
of the facility, we need to propose significant projects
which are worth for the investment.
As shown in Fig. \ref{fig:ub-db}, J-PARC could measure
structure functions in the medium-$x$ region $0.2<x<0.7$,
whereas RHIC probes a smaller-$x$ region ($x \sim 0.1$).
In order to determine all the partonic contributions 
to the nucleon spin, the polarized PDFs need to be
precisely understood from small $x$ to large $x$,
namely from the RHIC region to the J-PARC one.
Therefore, it is a complementary facility to RHIC and
other high-energy accelerators.
Although longitudinally-polarized PDFs become clearer
recently by various hadron and lepton reactions as shown 
in Fig. \ref{fig:polpdfs}, their flavor decomposition
and gluon distribution are not obvious yet.
If the proton beam is polarized with the designed energy
50 GeV, the J-PARC facility could significantly contribute
to the clarification of the origin of the nucleon spin.

In addition, there is new transverse spin physics,
such as twist-two transversity distributions,
by measuring double spin asymmetries of Drell-Yan processes
with transversely polarized protons.
The quark transversity distributions are unique in the sense
that they do not couple to the gluon polarization
due to their chiral-odd property. They are very 
different from the longitudinally-polarized PDFs.
The transversity distributions can be measured 
at medium $x$ with the polarized proton beam at J-PARC.

\vspace{-0.20cm}
\section{Summary}
\label{summary}

A wide range of spin projects are possible in particle and nuclear
physics at J-PARC. As the particle-physics topics, 
we introduced muon $g-2$, muon and neutron electric dipole moments, 
and time-reversal violation experiment in a kaon decay.
They could probe physics beyond the standard model by
precision measurements with the high-intensity advantage
of the J-PARC facility, and the measurements should shed light 
on new physics direction.
In hadron spin physics, the studies of GPDs and TMDs 
should clarify the origin of the nucleon spin and
also the three-dimensional structure of the nucleon
including the transverse structure.
In particular, the high-momentum beamline will be ready soon,
and it can be used for various topics on high-energy 
hadron spin physics.

\vspace{-0.20cm}
\section*{Acknowledgments}
This work was partially 
supported by the MEXT KAKENHI Grant Number 25105010.

\vspace{-0.20cm}


\end{document}